\documentclass[12pt,a4paper]{article}
\usepackage{amsfonts,latexsym}
\usepackage{graphicx,color}
\usepackage{dcolumn}
\usepackage{graphicx}
\usepackage{amsmath}
\usepackage{amsfonts}
\usepackage{amssymb}

\renewcommand{\thefootnote}{\fnsymbol{footnote}}

\begin{document}

\vspace{12mm}

\begin{center}
{{{\Large {\bf Superradiant instability of dyonic Reissner-Nordstr\"{o}m black holes  }}}}\\[10mm]

Yun Soo Myung$^{a,b}$\footnote{e-mail address: ysmyung@inje.ac.kr}\\[8mm]

{${}^a$Institute of Basic Sciences and Department  of Computer Simulation,  Inje University Gimhae 50834, Korea\\[0pt]}

{${}^b$Asia Pacific Center for Theoretical Physics, Pohang 37673, Korea}

\end{center}
\vspace{2mm}

\begin{abstract}
We investigate the superradiant instability of dyonic Reissner-Nordstr\"{o}m (dRN) black holes with two charges under an electrically charged massive scalar perturbation.
Two conditions for possessing a trapping well are obtained from analyzing asymptotic scalar potential and far-region wave functions.
It is clear that the superradiant instability is  not  allowed in the dRN black holes because the conditions for a trapping well are  not compatible with the superradiance condition.

\end{abstract}
\vspace{5mm}

\vspace{1.5cm}

\hspace{11.5cm}
\newpage
\renewcommand{\thefootnote}{\arabic{footnote}}
\setcounter{footnote}{0}


\section{Introduction}
It was first mentioned  that superradiant instability of  rotating  black hole with mass $M$ and angular momentum $J$ could be a continuous source of gravitational waves~\cite{Arvanitaki:2009fg}.
Ultralight bosons (axions or string axiverse) with mass $ \mu  \ll1/M$ could trigger superradiant instability of  the rotating black holes and form bosonic clouds which emit gravitational waves. These waves might  be detected by gravitational wave observations~\cite{Brito:2020lup,Yuan:2021ebu}.
Superradiant instability of astrophysical black holes has been used to constrain  physical parameters because ultralight bosons  might be
a promising candidate for the dark matter and it is not easy to detect its evidence by collision experiment in accelerator~\cite{Tsukada:2020lgt,Brito:2017wnc}.

On the other hand, superradiance may represent an amplification  of a charged massive scalar impinging on a static Reissner-Nordstr\"{o}m (RN) black hole with mass $M$ and charge $Q$ when  frequency $\omega$ and  charge $q$ of the scalar   obey the superradiance  condition $ \omega<\omega_c=q \Phi_H$ ($\Phi_H=qQ/r_+$, electric potential at horizon)~\cite{Bekenstein:1973mi}.
 We point out  that the literatures~\cite{Furuhashi:2004jk,Hod:2012wmy,Hod:2013nn} have used a shortened potential to show that  the gravitational attraction between RN black hole and a charged massive scalar cannot provide a confinement mechanism, triggering the superradiant instability. Especially, it was shown in~\cite{Furuhashi:2004jk} that
 a lower bound of `$M\mu>qQ$' has been proposed for a necessary condition to get a trapping well, but this condition is not satisfied simultaneously by  superradiant states.  Furthermore, one would say that  this inequality implies  the Newton-Coulomb requirement for the gravitational force to exceed the electrostatic force.
Recently, it was reported that one condition for no trapping well is  given by an upper bound of `$\mu M<qQ$'  in the Kerr-Newman  black hole whose non-rotating limit is the RN black hole~\cite{Xu:2020fgq}.
We would like to mention that  a correct potential was not used to derive this bound~\cite{Myung:2022kex}.

Some  aspects of the superradiance has been investigated in the RN black hole~\cite{DiMenza:2014vpa} and the absorption cross section of a charged massive scalar~\cite{Benone:2015bst} was obtained in the RN black hole.
It is important to note  that the superradiant instability is hard to arise naturally  from a charged  massive scalar propagation around the RN black holes.
However, the superradiant instability of a charged massive scalar  could be obtained in certain parameter space if a cavity is placed around the RN black hole~\cite{Herdeiro:2013pia,Degollado:2013bha,Hod:2013fvl,Hod:2016kpm}. This is understood  as the charged black hole-mirror bomb.
Here, numerical techniques have used to show the lower bound on $q$ ($q>\mu$), which may be  considered as  a necessary condition for the superradiant instability when a cavity is placed around the RN black hole~\cite{Herdeiro:2013pia}.

In addition, a magnetic field surrounding a black hole could provide a confinement mechanism because it induces an effective mass $\mu_{\rm eff}\propto B$~\cite{Konoplya:2008hj,Konoplya:2007yy}. This implies that  a magnetic field  has an effect on the superradiance of a black hole, which  may have astrophysical applications. Investigating  a massless scalar perturbation in a magnetized Schwarzschild  black hole with magetic field $B$  leads to triggering the superradiant instability in a confining box of size $1/B$~\cite{Brito:2014nja}.  Here, $Bm$ plays a role of mass $\mu_{\rm eff}$ as $(Bm)^2$-term in the potential $V_{\rm eff}(r,\theta)$.

In this connection, the superradiant stable regime for  dyonic  Reissner-Nordstr\"{o}m (dRN) black holes with electric $Q_e$ and magnetic charge $Q_m$  was introduced to obtain  an upper bound of $(Q_e^2+Q_m^2)/M^2<24/25$ by analyzing an effective potential $V(r)$ in~\cite{Zou:2021mwa}. We note that this black hole becomes  an asymptotically flat RN black hole when replacing $Q_e^2+Q_m^2$ by $Q^2$, but it was suggested  that the dRN black hole spacetime could provide a magnetic field.   However, we wish to point out that a correct potential $U_1(r)$ was not used  to derive this bound when considering an electrically charged massive scalar perturbation. Therefore, this superradiant stable regime is not regarded as a correct bound.

In this work, wish to investigate  the superradiant (un)stable regime for dRN black holes by employing the correct scalar potential $V_{\rm dRN}(r)=U_1(r)$ which differs by $q^2Q_m^2/r^2$ from the scalar potential $V_{\rm RN}(r)$ around the RN black hole. It is important to clarify that an inclusion of  a magnetic charge $Q_m$ in $q^2Q_m^2/r^2$  could not induce a magnetic field like $\mu_{\rm eff} \propto B$ in the scalar potential $V_{\rm dRN}(r)$. This is likely to upset  the original motivation of introducing dRN black holes to get a magnetic filed. We will obtain two conditions for getting a trapping well by analyzing all asymptotic potentials and far-region scalar wave functions.
These conditions are given by $M\mu^2>qQ_e\omega~(M\mu/qQ_e>\omega/\mu)$ and $a<0~[B/(2A)>k+1/2]$. The former condition indicates the Newton-Coulomb requirement for the gravitational force to exceed the electrostatic force, while the latter represents the condition to get a quasibound state.
On the other hand, two conditions for no trapping well are $M\mu^2<qQ_e\omega~(M\mu/qQ_e<\omega/\mu<1)$ and $a>0~[B/(2A)<k+1/2]$. The former  represents   the Coulomb-Newton requirement for the electrostatic  force to exceed the gravitational force, while the latter denotes the condition to get  a bound state.
Furthermore, we explore  how the known  bounds of $M\mu>qQ$ ($M\mu<qQ$) arise  from the pseudo-Newtonian  potential for a charged particle with mass $\mu$ and charge $q$ moving around the RN black holes with mass $M$ and charge $Q$~\cite{Ivanov:2005cf}.

\section{Propagation of scalar  on the dRN black holes }
The dRN black hole  takes the form~\cite{Chen:2012pt,Zou:2021mwa}
\begin{eqnarray}
ds_{\rm dRN}^2&=&\bar{g}_{\mu\nu}dx^{\mu}dx^{\nu} \nonumber \\
&=& -f_{\rm dRN}(r)dt^2+\frac{dr^2}{f_{\rm dRN}(r)}
+r^2 (d\theta^2 +\sin^2 \theta d\phi^2)   \label{d-RN}
\end{eqnarray}
with the dRN metric function and its electromagnetic potential
\begin{equation}
f_{\rm dRN}(r)=1-\frac{2M}{r}+\frac{Q_e^2+Q_m^2}{r^2},~\bar{A}_\mu=\Big[-\frac{Q_e}{r},0,0, Q_m(\cos\theta \mp 1)\Big],
\end{equation}
where the upper minus (lower plus) denotes the north-half sphere (south-half sphere) of the dRN black hole.
Here,  the line element Eq.(\ref{d-RN}) is stationary and  spherically  symmetric.
In this case, the outer and inner horizons are   given by
\begin{equation}
r_\pm=M\pm\sqrt{M^2-Q_e^2-Q_m^2},
\end{equation}
where  the condition $(Q^2_e+Q^2_m)/M^2\le 1$ is necessary for the existence of a horizon including an extremal black hole. For $Q_m=0$, one finds the RN black hole with mass $M$ and charge $Q_e$. Also,  if one replaces $Q_e^2+Q_m^2$ by $Q^2$, one finds the RN black hole with $f_{\rm RN}(r)=1-2M/r+Q^2/r^2$.

An electrically  charged massive scalar perturbation $\Phi$  on the background of dRN black holes is described  by
\begin{equation}
(\bar{\nabla}^\mu-i q \bar{A}^\mu)(\bar{\nabla}_\mu-i q \bar{A}_\mu)^*\Phi-\mu^2\Phi=0,\label{phi-eq1}
\end{equation}
where $q$ denotes an electric charge  of  $\Phi$ and $\mu$ represents a mass of  $\Phi$.

Taking into account the
background (\ref{d-RN}), it is useful to separate the scalar perturbation
into modes
\begin{equation}
\Phi(t,r,\theta,\phi)=e^{-i\omega t+im\phi} Y(\theta) R(r), \label{sep}
\end{equation}
where $Y(\theta)$ denotes the angular part of solution and $R(r)$ represents the radial part.
Considering the north half-sphere ($0\le \theta \le \pi/2$)  and south-half sphere ($\pi/2\le \theta \le \pi$) of the dRN black hole separately, one finds their equations
\begin{eqnarray}
 \frac{1}{\sin \theta}\partial_{\theta}\Big(
\sin \theta
\partial_{\theta} Y_1(\theta) \Big )&+& \left [\lambda_1-\frac{m^2+2(mqQ_m+q^2Q_m^2)(1-\cos\theta)}{\sin ^2{\theta}} \right ]Y_1(\theta) =0, \label{wave-ang1}\\
\frac{1}{\sin \theta}\partial_{\theta}\Big(
\sin \theta
\partial_{\theta} Y_2(\theta) \Big )&+& \left [\lambda_2-\frac{m^2-2(mqQ_m+q^2Q_m^2)(1+\cos\theta)}{\sin ^2{\theta}} \right ]Y_2(\theta) =0.\label{wave-ang2}
\end{eqnarray}
Requiring that $\{Y_1(\chi),Y_2(\eta)\}$ with $\chi,\eta=\cos\theta$ be finite at north and south poles, the solutions to Eqs.(\ref{wave-ang1}) and (\ref{wave-ang2}) are given by
\begin{eqnarray}
Y_1(\chi)&=&c_1(1-\chi)^{\frac{m}{2}}(1+\chi)^{-\frac{m+2qQ_m}{2}}{}_2F_1\Big[\alpha_1,\beta_1,1+m;\frac{1-\chi}{2}\Big], \\
Y_2(\eta)&=&c_2(1-\eta)^{-\frac{m-2qQ_m}{2}}(1+\eta)^{\frac{m}{2}}{}_2F_1\Big[\alpha_2,\beta_2,1-m+2qQ_m;\frac{1-\eta}{2}\Big]
\end{eqnarray}
with the hypergeometric functions ${}_2F_1[\cdots]$ and their coefficients
\begin{equation}
\alpha_1/\beta_1=\frac{1-2qQ_m\pm\sqrt{1+4\lambda_1}}{2},~\alpha_2/\beta_2=\frac{1+2qQ_m\pm\sqrt{1+4\lambda_2}}{2}.
\end{equation}
Furthermore, imposing the finiteness of $(1-\chi)^{m/2}$ and $(1+\eta)^{m/2}$ leads to $m\ge 0$. Also,
considering the convergence of two ${}_2F_1[\cdots]$ under $|(1-\chi)/2|\le 1$ and $|(1-\eta)/2|\le 1$, one finds  the  quantization condition for magnetic charge,  the constraint on $\lambda_1$ and $\lambda_2$, and the upper bound on the magnetic charge, respectively,
\begin{equation}
qQ_m=~{\rm integer}, \quad  \lambda_1=\lambda_2=l(l+1),\quad q^2Q_m^2 < l(l+1).
\end{equation}
On the other hand,  the radial  equation for $R_l(r)$ with $\Delta=r^2f_{\rm dRN}(r)$ is given by
\begin{eqnarray}
\Delta \partial_r \Big( \Delta \partial_r R_{\ell}(r) \Big)+\tilde{U}(r)R_{l}(r)=0
\label{wave-rad}
\end{eqnarray}
with
\begin{eqnarray}
\tilde{U}(r)=(\omega r^2-qQ_er)^2-\Delta[\mu^2r^2 +l(l+1)-q^2Q_m^2]. \label{u-pot}
\end{eqnarray}
Here, the last term of $q^2Q_m^2$ is an additional contribution from the angular part.
It is worth noting that Eq.(\ref{wave-rad}) is usually  used  to obtain  exact solutions.
In this direction, the radial solution to the Klein-Gordon equation for a massive charged scalar propagating around the Kerr-Newman black hole
is given by the confluent Heun function HeunC($\alpha,\pm\beta,\gamma,\delta,\eta;z)$~\cite{Bezerra:2013iha,Kraniotis:2016maw}.
Two near-horizon and asymptotic  solutions derived from the confluent Heun function were compared with the Whittaker's form  which is obtained asymptotically from  
the confluent Kummer and Tricomi functions as well as in the near-horizon.

Now, introducing a tortoise coordinate $r_*$ defined by
\begin{equation}
r_*=\int\frac{dr}{f_{\rm dRN}(r)}=r+\frac{r_+^2}{r_+-r_-}\ln(r-r_+)-\frac{r_-^2}{r_+-r_-}\ln(r-r_-),
\end{equation}
one may derive the Schr\"{o}dinger-type equation when setting $\Psi_{l}(r)=r R_{l}(r)$ from  Eq.(\ref{wave-rad})
\begin{equation}
\frac{d^2\Psi_{l}(r_*)}{dr_*^2}+\Big[\omega^2-V_{\rm dRN}(r)\Big]\Psi_{l}(r_*)=0, \label{sch-eq}
\end{equation}
where the potential  $V_{\rm dRN}(r)$ is found to be
\begin{eqnarray}
V_{\rm dRN}(r)=\omega^2-\Big(\omega-\frac{qQ_e}{r}\Big)^2+f_{\rm dRN}(r)\Big[\mu^2+\frac{l(l+1)}{r^2}-\frac{q^2Q_m^2}{r^2}+\frac{2(Mr-Q_e^2-Q_m^2)}{r^4}\Big].\label{e-pot}
\end{eqnarray}
Here, the last term in $[\cdots]$ comes from the introduction of the tortoise coordinate $r_*$. We wish to point out asymmetric roles of two charges $Q_e$ and $Q_m$.
We observe that an electric charge $Q_e$ appears in the second term because an electrically charged massive scalar perturbation $\Phi$ is introduced as a probe field.
On the other hand, a magnetic charge $Q_m$ appears as the third term in $[\cdots]$. Comparing it with the mass term $\mu^2$  leads to the fact that an inclusion of $Q_m$
does not induce  a mass term. It may play the role of a centripetal term when comparing with a centrifugal  $l(l+1)/r^2$-term.
Thus, it is interesting to mention that $V_{\rm dRN}(r)$ differs by the third term from the potential for an electrically charged massive  scalar in the RN black hole background with mass $M$  and charge $Q$ as~\cite{DiMenza:2014vpa,Benone:2015bst,Herdeiro:2013pia,Degollado:2013bha,Hod:2013fvl,Hod:2016kpm}
 \begin{eqnarray}
V_{\rm RN}(r)=\omega^2-\Big(\omega-\frac{qQ}{r}\Big)^2+f_{\rm RN}(r)\Big[\mu^2+\frac{l(l+1)}{r^2}+\frac{2(Mr-Q^2)}{r^4}\Big].\label{RN-pot}
\end{eqnarray}

Taking the asymptotic form of Eq.(\ref{sch-eq}) and its near-horizon form, one finds  plane-wave  solutions
\begin{eqnarray}
\Psi^{\infty}(r_*)&\sim&  e^{-i\sqrt{\omega^2-\mu^2} r_*}(\leftarrow)+{\cal R}e^{+i\sqrt{\omega^2-\mu^2} r_*}(\rightarrow),\quad r_*\to +\infty(r\to \infty) , \label{asymp1}\\
\Psi^{-\infty}(r_*)&\sim& {\cal T} e^{-i(\omega-\omega_c) r_*}(\leftarrow),\quad r_*\to -\infty(r\to r_+), \label{asymp2}
\end{eqnarray}
where  ${\cal T}({\cal R})$ are  the transmission (reflection) amplitudes.
Here, we may have  $V(r\to r_+)=\omega^2-(\omega-\omega_c)^2$ with $\omega_c=qQ_e/r_+$ in the near-horizon limit.
 Imposing the flux conservation, we obtain a relation between reflection and transmission coefficients as
\begin{equation}
|{\cal R}|^2=1-\frac{\omega-\omega_c}{\sqrt{\omega^2-\mu^2}}|{\cal T}|^2,
\end{equation}
which  means that the scalar waves with $\omega>\mu$ propagate to infinity and the superradiant scattering  may occur ($\rightarrow,~|{\cal R}|^2>|{\cal I}|^2$) when $\omega<\omega_c$ (superradiance condition) is satisfied.

On the other hand, one may choose the bound state condition ($\omega<\mu$) to have an exponentially decaying scalar as it tends to zero  at infinity
 \begin{equation}
 \Psi^{\rm b,\infty}(r) \sim e^{-\sqrt{\mu^2-\omega^2} r} \rightarrow 0. \label{asymp-b}
 \end{equation}
Solving Eq.(\ref{wave-rad}) directly, the frequency $\omega $ is permitted to be complex (small complex modification)
 as
\begin{equation}
\omega=\omega_{\rm R}+i \omega_{\rm I}.
\end{equation}
Here the sign of $\omega_{\rm I}$ determines the solution which  is decaying ($\omega_{\rm I}<0$) or growing ($\omega_{\rm I}>0$) in time.

Finally, we  mention   two cases  for a electrically charged massive  scalar propagating around the dRN black holes according to the potential analysis.\\
Case (i) superradiant stability: $\omega<\omega_c$ and   $\omega<\mu$ without a positive trapping well. \\
Case (ii) superradiant instability: $\omega<\omega_c$ and   $\omega<\mu$  with  a positive trapping well.\\
Solving the radial equation (\ref{wave-rad}) directly leads to the real  frequency ($\omega_{\rm R}$) and the imaginary one ($\omega_{\rm I}$). In this case, one could describe the above cases again: \\
Case (i): $\omega_{\rm I}<0$ and  $\omega_{\rm R}<\omega_c$. The  solution is stable (decaying in time). \\
Case (ii): $\omega_{\rm I}>0$ and   $\omega_{\rm R}<\omega_c$. The  solution is unstable (growing in time).

\section{Potential analysis}
First of all, we start with  an explicit form of the scalar potential (\ref{e-pot}) as
\begin{eqnarray}
\tilde{V}_{\rm dRN}(r)&=& \mu^2-\frac{2(M\mu^2-qQ_e\omega)}{r} +\frac{l(l+1)+(Q_e^2+Q_m^2)(\mu^2-q^2)}{r^2}\nonumber \\
&-&\frac{2M[l(l+1)-q^2Q_m^2-1]}{r^3}+\frac{(Q_e^2+Q_m^2)[l(l+1)-q^2Q_m^2-2]-4M^2}{r^4}\nonumber \\
    &+&\frac{6M (Q_e^2+Q_m^2)}{r^5}-\frac{2(Q_e^2+Q_m^2)^2}{r^6}, \label{til-pot}
\end{eqnarray}
where the first line is generated from $\tilde{U}(r)$ in the radial equation (\ref{u-pot}) solely, the second one comes from both  $\tilde{U}(r)$ and the introduction of tortoise coordinate $r_*$,
and the third line is generated from  introducing tortoise coordinate $r_*$. It is important to mention that  the potential in the first line plays an essential role in analyzing the superradiant (in)stability.

An asymptotic form of the potential is given by
\begin{equation}
V_{\rm aadR}(r)=\mu^2-\frac{2(M\mu^2-qQ_e\omega)}{r}\equiv 1-\frac{B}{r}, \label{asym-p}
\end{equation}
which is obtained in the asymptotic region of $r\to \infty$ from (\ref{til-pot}). Here, we note the absence of magnetic charge $Q_m$.
 This Newton-Coulomb potential could be used to find the condition to get a trapping well as
\begin{equation}
V'_{\rm aadR}(r)>0 \to \quad M\mu^2>qQ_e \omega~(B>0). \label{cond-t}
\end{equation}
This  indicates the Newton-Coulomb requirement for the gravitational (attractive) force to exceed the electrostatic (repulsive) force.
On the other hand,  the condition for no trapping well is realized as
\begin{equation}
V'_{\rm aadR}(r)<0 \to \quad M\mu^2<qQ_e \omega~(B<0), \label{nocond-t}
\end{equation}
which  describes the Coulomb-Newton requirement for the electrostatic  force to exceed the gravitational force.
The case of $V'_{\rm aadR}(r)=0$ corresponds to the absence of the  Newton-Coulomb-term ($0/r$).

However, Eq.(\ref{cond-t}) [Eq.(\ref{nocond-t})] is not a sufficient condition for determining a trapping well [no trapping well].
This is so because a case of   $V'_{\rm aadR}(r)>0$ includes three potential types: potential with a trapping well (local maximum), potential with a tiny well, and an increasing potential without any extrema.  Also, the case of $V'_{\rm aadR}(r)<0$ implies  either a potential without trapping well or a potential with a local maximum.
Therefore, we have to find the other conditions to specify a trapping well. For this purpose,
we are necessary to introduce  the far-region potential appeared in the large $r$ region from  (\ref{til-pot}) as
\begin{eqnarray}
V_{\rm adR}(r)&=&\mu^2-\frac{2(M\mu^2-qQ_e\omega)}{r}+\frac{l(l+1)+(Q_e^2+Q_m^2)(\mu^2-q^2)}{r^2} \label{fr-p} \\
&\equiv& \mu^2-\frac{B}{r}+\frac{C}{r^2}, \label{fr-pd}
\end{eqnarray}
where  $C$-term plays a crucial  role in making a trapping well. We observe from $V_{\rm adR}(r)$ that an inclusion of  magnetic charge $Q_m$ plays an equal role of the electric charge $Q_e$, but it does not play the role of a magnetic field like $\mu \propto B$. It is worth noting that  two charges of the dRN black holes induce repulsive (attractive) effects to the scalar potential for $\mu>q~(\mu<q)$. However, such effects are limited because the condition for the existence of a horizon is given by $(Q_e^2+Q_m^2)/M^2\le 1$.  This can be  seen easily from  a centrifugal potential ($l(l+1)/r^2$-term)  which may have  a greatly repulsive effect on making a trapping well for large $l$.  Actually, there is no way to make a trapping well (local minimum)  if one keeps the Newton-Coulomb potential  $V_{\rm aadR}(r)$ only.
Here, we  have an extremal point ($r_{\rm e}$)
\begin{equation}
V'_{\rm adR}(r_{\rm e})=0 \quad \to r_{\rm e}=\frac{l(l+1) +(Q_e^2+Q_m^2)(\mu^2-q^2)}{M\mu^2-qQ_e\omega}=\frac{2C}{B}>r_+,
\end{equation}
which becomes either a local minimum or a local maximum, located far from the outer horizon.
\begin{figure*}[t!]
   \centering
  \includegraphics{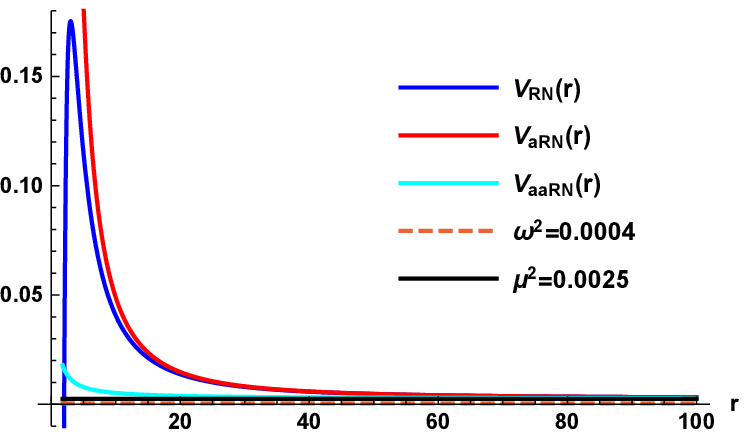}
   \hfill%
  \includegraphics{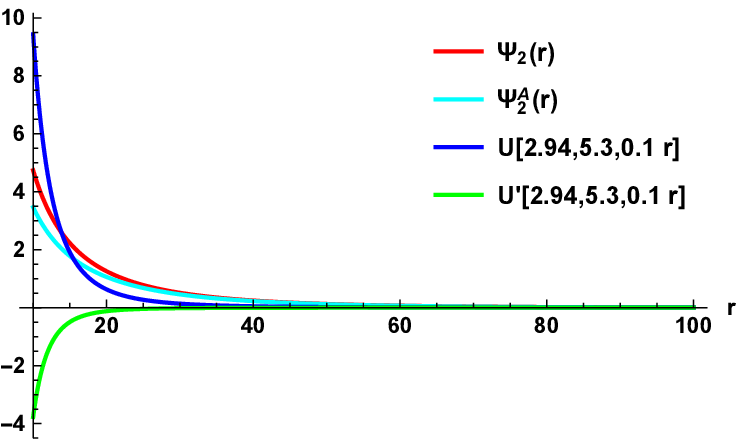}
\caption{(Left) Potential $V_{\rm dRN}(r)$   as function of $r\in[r_+ =1.77,100]$ with $M=1,Q_e=0.4,Q_m=0.5,\omega=0.02,q=2,l=2,\mu=0.05$. We have no trapping well here.
We note   $\omega<\mu$ and $\omega<\omega_c(=0.45)$  to meet a superradiant stability with $V'_{\rm aadR}(r)<0$.  (Right) Far-region scalar function $\Psi_{2}(r)$ and its asymptotic form  $\Psi^{\rm A}_{2}(r)$ with confluent hypergeometric  $U[2.94,5.3,0.1r]$  represent bound states.  }
\end{figure*}
\begin{figure*}[t!]
   \centering
  \includegraphics{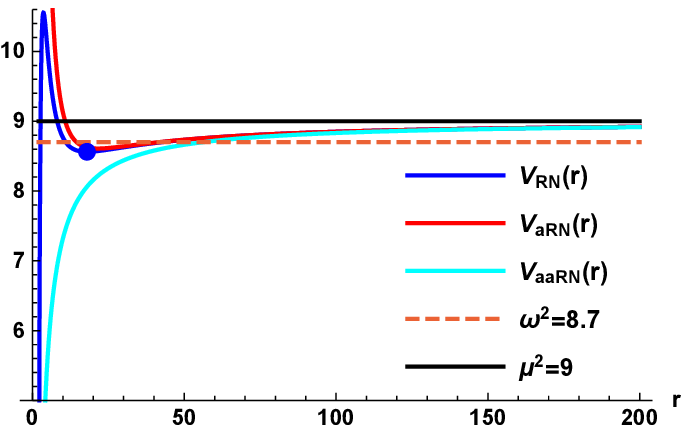}
   \hfill%
  \includegraphics{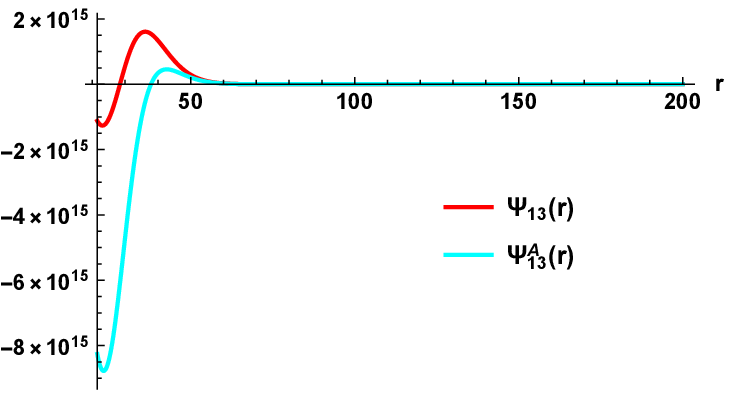}
\caption{(Left) Potential $V_{\rm dRN}(r)$   as function of $r\in[r_+ =1.99,200]$ with $M=1,Q_e=0.01,Q_m=0.05,\omega=2.95,q=20,l=13,\mu=3$.
A blue dot ($\bullet$) denotes a local minimum at $r=17.9$, implying a trapping well. Here, one has $V'_{\rm aadR}(r)>0$.
We note  that $\omega_c(=0.1)<\omega$  does not satisfy  the superradiance condition.  (Right) Far-region scalar function $\Psi_{13}(r)$ and its asymptotic form  $\Psi^{\rm A}_{13}(r)$ as functions of $r\in[21.5,200]$ represent quasibound states. }
\end{figure*}
\begin{figure*}[t!]
   \centering
  \includegraphics{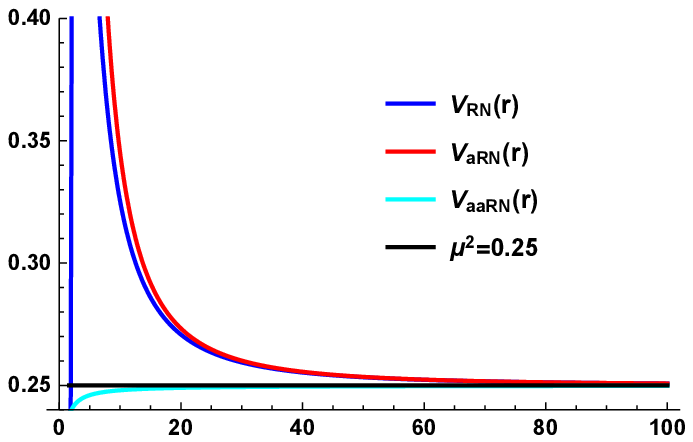}
   \hfill%
  \includegraphics{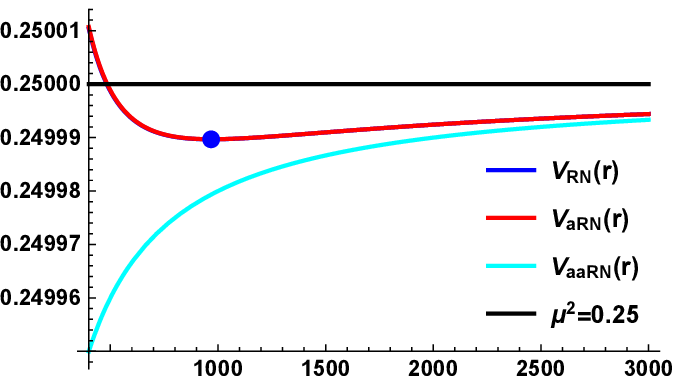}
\caption{(Left)Potential $V_{\rm dRN}(r)$   as function of $r\in[r_+ =1.62,100]$ with $M=1,Q_e=0.6,Q_m=0.5,\omega=0.2,q=2,l=3,\mu=0.5$.
We note   $\omega<\mu$ and $\omega<\omega_c(=0.74)$  to meet a superradiant stability with $V'_{\rm aadR}(r)>0$. Here, no trapping well is found apparently.  (Right) $V_{\rm dRN}(r)\simeq V_{\rm adR}(r)$ have  a tiny well ($\bullet$) located at $r=971$. $V_{\rm aadR}(r)$ approaches them for $r>971$.  }
\end{figure*}

Hereafter, we choose $M=1$  such that $M\mu$ becomes $\mu$  for  simplicity.
We wish to introduce five distinct potentials to test their presence of a trapping well. We expect that these  potentials offer a complete classification of all potentials.
First of all, we consider  a potential appeared in (Left) Fig. 1. It includes no  trapping well,  being consistent with $V'_{\rm aadR}(r)<0$.
It is curious to note that (Left) Fig. 2 corresponds to a potential with a trapping well ($\bullet$), implying $V'_{\rm aadR}(r)>0$.
On the other hand, it is interesting to explore a secret nature of a potential of (Left) Fig. 3.
In this case, we observe apparently  that there is no trapping wells. But, the condition of $V'_{\rm aadR}(r)>0$  may imply a trapping well. So, it seems that $V'_{\rm aadR}(r)>0$  is not compatible with our expectation of no trapping well.
To resolve it, we  note  that a tiny  well ($\bullet$) in (Right) Fig. 3  is located at a very large distance of $r=971$ in $V_{\rm dRN}(r)\simeq V_{\rm adR}(r)$, but its presence does not affect the superadiant stability.  This means that $V'_{\rm aadR}(r)>0$ implies either a trapping well or a tiny well.  Hence, it is necessary to find a further condition for a trapping well in the next section.

Interestingly,  (Left) Fig. 5 shows a different potential with $V'_{\rm aadR}(r)<0$, but it has a local maximum ($\bullet$) located at $r=142$ in $V_{\rm dRN}(r)\simeq V_{\rm adR}(r)$ [see (Right) Fig. 5]. It looks like an upside-down figure of (Right) Fig. 3.
A final potential with $V'_{\rm aadR}(r)>0$ appeared in (Left) Fig. 7 represents an increasing function approaching $\mu^2=32.5$, implying no trapping well. Here, $V_{\rm adR}(r)$ includes a well inside the outer horizon, becoming a meaningless case. This potential may correspond to a boundary between potential with trapping well and potential without trapping well.
\begin{figure*}[t!]
   \centering
  \includegraphics{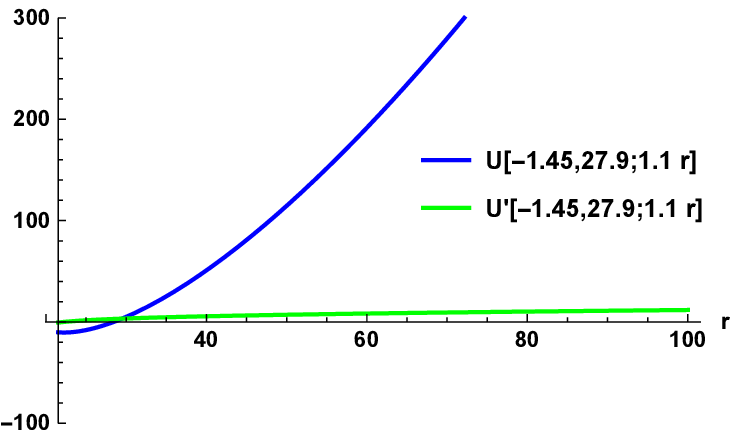}
   \hfill%
  \includegraphics{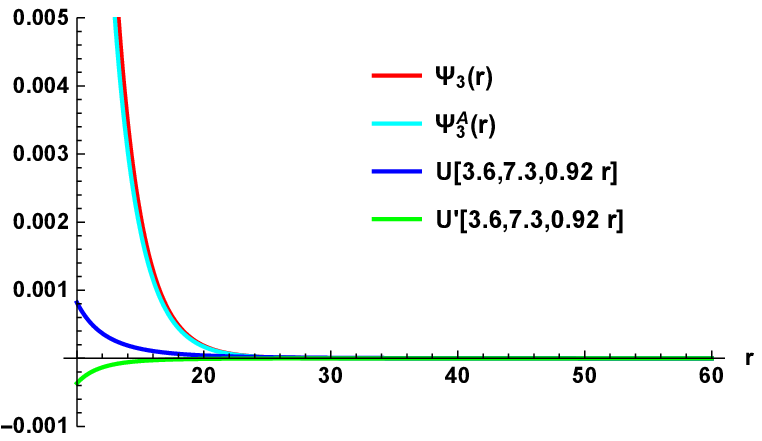}
\caption{(Left) Hypergeometric function $U[-1.45,27.9,1.1r]$ and its first derivative $U'[-1.45,27.9,1.1r]$ represent (Left) Fig. 2.   (Right) Far-region scalar function $\Psi_{3}(r)$ and its asymptotic form  $\Psi^{\rm A}_{3}(r)\simeq \Psi_{3}(r)$ with $U[3.6,7.3,0.92r]$  represent bound states appeared in (Left) Fig. 3. }
\end{figure*}
\begin{figure*}[t!]
   \centering
  \includegraphics{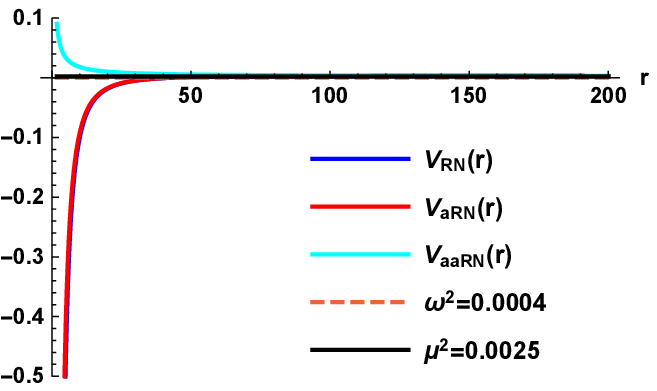}
  \hfill%
  \includegraphics{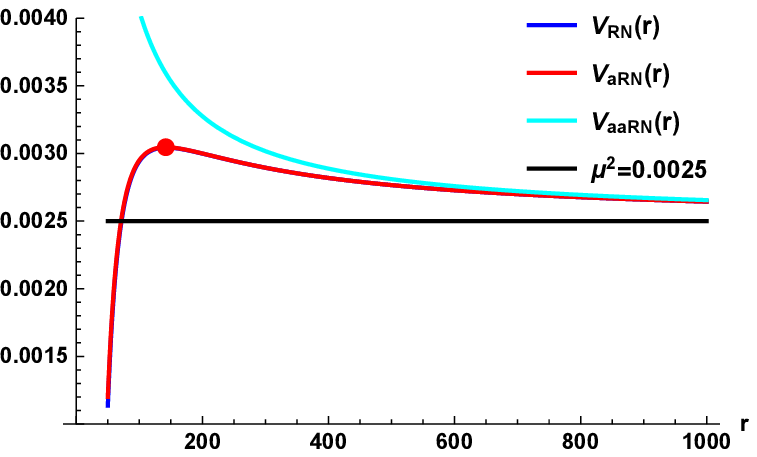}
\caption{ Local maximum potential $V_{\rm dRN}(r)\simeq V_{\rm adR}(r)$   as function of $r\in[r_+ =1.77,100]$ with $M=1,Q_e=0.4,Q_m=0.5,\omega=0.02,q=10,l=5,\mu=0.05$.
We note   $\omega<\mu$ and $\omega<\omega_c(=2.26)$  with $V'_{\rm aadR}(r)<0$.  (Right) $V_{\rm dRN}(r)\simeq V_{\rm adR}(r)$ have  a local maximum ($\bullet$) located at $r=142$. $V_{\rm aadR}(r)$ approaches them for $r>142$.  }
\end{figure*}
 \begin{figure*}[t!]
   \centering
  \includegraphics{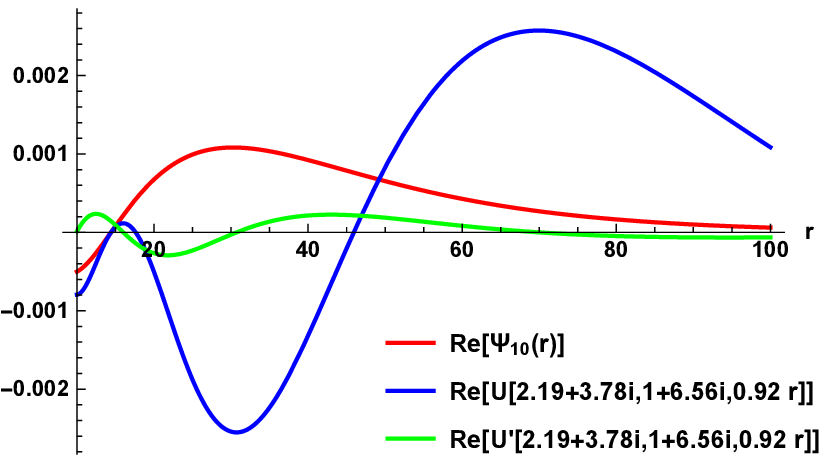}
  \hfill%
  \includegraphics{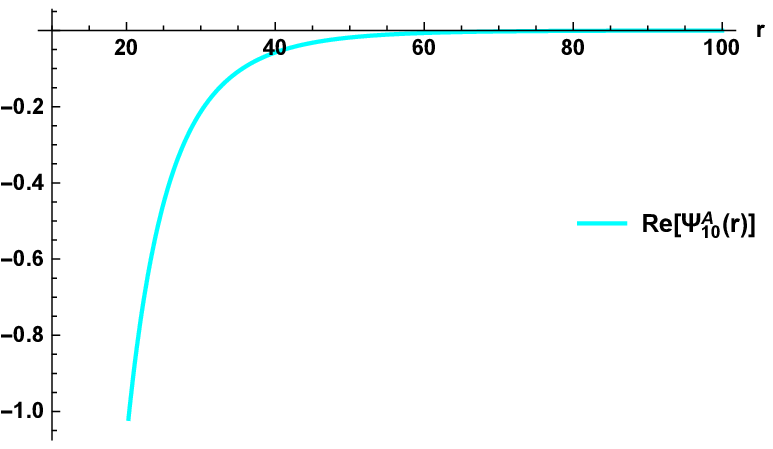}
\caption{ (Left) Oscillating function Re[$\Psi_{10}(r)$] with  Re[$U[2.19+3.78i,1+6.56i;0.92r]]$ and Re[$U'[2.19+3.78i,1+6.56i;0.92r]$] for (Left) Fig. 5. (Right) Its asymptotic wave function Re[$\Psi^{\rm A}_{10}(r)$] represents a negative bound state.}
\end{figure*}
\begin{figure*}[t!]
   \centering
  \includegraphics{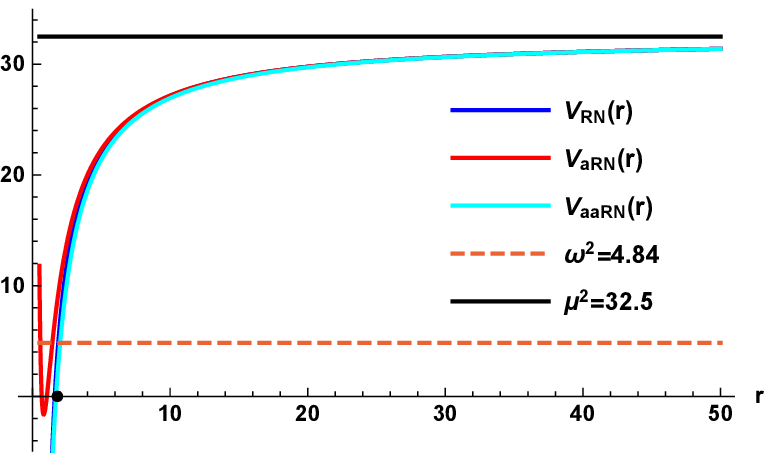}
  \hfill%
  \includegraphics{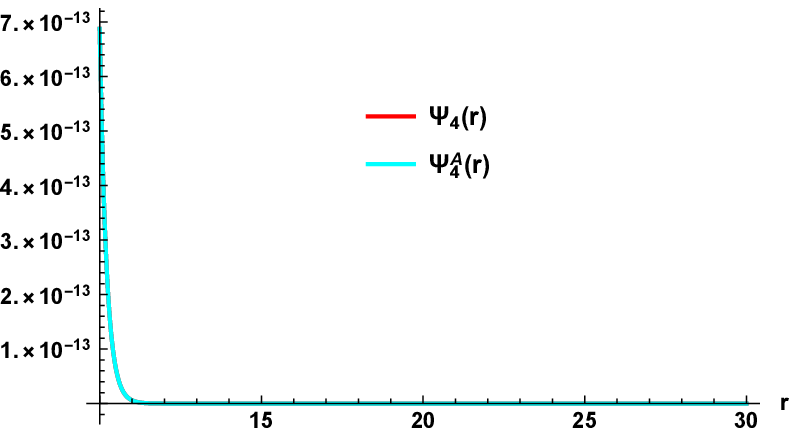}
\caption{(Left) Increasing potential $V_{\rm dRN}(r)\simeq V_{\rm adR}(r)\simeq V_{\rm aadR}(r)$   as function of $r\in[r_+ =1.81,50]$ with $M=1,Q_e=0.43,Q_m=0.4,\omega=2.2,q=5,l=4,\mu=5.7$.
A local minimum of $V_{\rm adR}(r)$ is found inside the outer horizon at $r=r_+$ ($\bullet$).
We note   $\omega<\omega_c(=4.75)$ and $\omega<\mu$   with $V'_{\rm aadR}(r)>0$.  (Right)  Far-region scalar  $\Psi_{4}(r)$ and its asymptotic form  $\Psi^{\rm A}_{4}(r)\simeq \Psi_{4}(r)$ represent a half of a peak. }
\end{figure*}
\begin{figure*}[t!]
   \centering
  \includegraphics{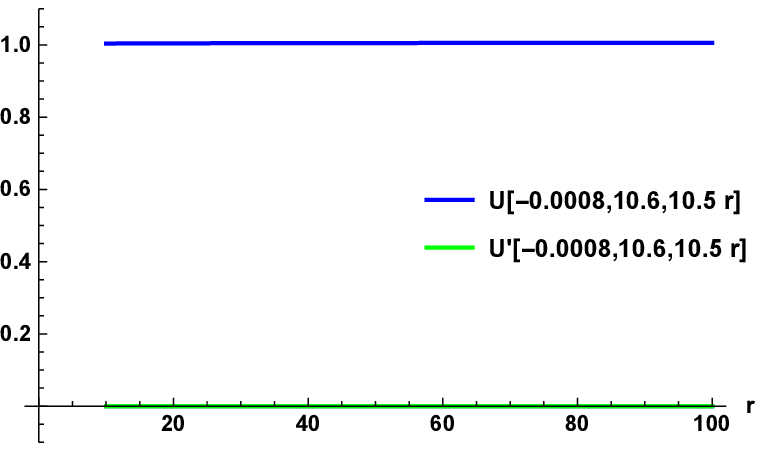}
\caption{(Left) Constant hypergeometric function $U[-0.0008,10.6,10.5 r]$ and its derivative $U'[-0.0008,10.6,10.5 r]\simeq 0$ for (Left) Fig. 7.}
\end{figure*}

\section{Far-region and asymptotic wave functions}

In the previous section, we are aware that the potential analysis is not enough to determine the presence of a trapping well because  $V'_{\rm aadR}(r)>0$ in Eq.(\ref{cond-t}) is not a sufficient condition to guarantee  a trapping well.
Therefore, it is necessary  to investigate  the scalar wave forms in the  far-region and asymptotic region by distinguishing between quasibound states (trapping well)  and  bound states (no trapping well).

In the far-region where one takes $r_*\simeq r$, we obtain an  equation from  (\ref{sch-eq}) together with (\ref{fr-p}) as
\begin{equation}
\Big[\frac{d^2}{dr^2}+\omega^2-V_{\rm adR}(r)\Big]\Psi_{l}(r)=0,
\end{equation}
which could be rewritten explicitly as
\begin{equation}
\Big[\frac{d^2}{dr^2}-A^2+\frac{B}{r}-\frac{C}{r^2}\Big]\Psi_{l}(r)=0 \label{p-whitt}
\end{equation}
with
\begin{equation}
A=\sqrt{\mu^2-\omega^2}.
\end{equation}
Here, the bound state condition $\omega<\mu$  always implies $A>0$.
Introducing $x=2Ar$, Eq.(\ref{p-whitt}) takes the Whittaker's equation as
\begin{equation}
\frac{d^2\Psi_{l}(x)}{dx^2} +\Big[-\frac{1}{4}+\frac{B}{2Ax}-\frac{C}{x^2}\Big]\Psi_{l}(x)=0. \label{whitt-eq}
\end{equation}
Its solution is given exactly by either the Whittaker function $W[a,b;cr]$~\cite{Whittaker} or the confluent  hypergeometric function $U(a,b;cr)$ as
\begin{eqnarray}
\Psi_{l}(r)&=&c_3W\Big[\frac{B}{2A},k;2Ar\Big] \nonumber \\
&=&c_3 e^{-Ar} \Big(2A r\Big)^{\frac{1}{2}+k}
  U\Big[k+\frac{1}{2}-\frac{B}{2A},1+2k;2A r\Big] \label{wavef-1}
\end{eqnarray}
with
\begin{equation}
k=\frac{1}{2}\sqrt{1+4C}.
\end{equation}
 Here, imposing  a real $k$ implies a condition for $C$
 \begin{equation}
 C>-\frac{1}{4}.
 \end{equation}
A large $r$-form of $U[a,b;cr]$ defined by
\begin{equation}
U[a,b;cr\to \infty]\rightarrow\quad  (cr)^{-a}\Big[1-\frac{a(1+a-b)}{cr} +{\cal O}\Big(\frac{1}{cr}\Big)^2\Big] \label{large-U}
\end{equation}
suggests   that one finds  a decreasing function $U[a,b;cr]$ for  $a>0$, whereas one has an increasing function $U[a,b;cr]$ for  $a<0$.
Plugging  Eq.(\ref{large-U}) into Eq.(\ref{wavef-1}) leads to an asymptotic wave function
as
\begin{equation}
\Psi^{\rm A}_{l}(r)\simeq e^{-Ar} \Big(2A r\Big)^{\frac{B}{2A}}\Big[1-\frac{(k+\frac{1}{2}-\frac{B}{2A})(-k+\frac{1}{2}-\frac{B}{2A})}{(2 Ar)}\Big]. \label{wavef-2}
\end{equation}
Here, we observe that taking into account  $e^{-Ar}$, a quasibound state may appear  for $\frac{B}{2A}>0$, whereas  a bound state may appear  for $\frac{B}{2A}<0$.
In addition, considering the first derivative $U'[a,b;cr]=-acU [1+a,1+b;cr]$ with respect to $r (c>0)$,
it implies  that
the condition for a trapping well is given by $
U'[a,b;cr]>0$, indicating $ a<0$.
On the other hand,  the condition for no trapping well takes the upper bound
$U'[a,b;cr]<0$,  showing  $ a>0$.

We are in a position to describe far-region and  asymptotic wave function according to the potential classification.
Considering the potential  without trapping well whose asymptotic derivative is negative ($V_{\rm aadR}'(r)<0$) [(Left) Fig. 1],  (Right) Fig.1 shows that all $\Psi_2(r),~\Psi_2^{\rm A}(r),$ and $U[2.94,5.3;0.1r]$  are decreasing functions with a negative $U'[2.94,5.3;0.1r]$. They  describe  bound state. This case represents no trapping well clearly.

Let us observe a radial mode  $\Psi_{13}(r)$ and its asymptotic mode $\Psi_{13}^{\rm A}(r)$ for the potential with a trapping well ($V_{\rm aadR}'(r)>0$) [see (Left) Fig. 2].
As is shown in (Right) Fig.2, they represent   quasi-bound states.
In this case, one has an increasing function $U[-2.35,27.9;1.1r]$  and its increasing derivative $U'[-2.35,27.9;1.1r]$ appeared in (Left) Fig. 4.
This case shows  a trapping well clearly.

Choosing a  potential [(Left) Fig. 3] without  trapping well, all $\Psi_3(r),~\Psi_3^{\rm A}(r),$ and $U[3.6,7.3;0.92r]$  are decreasing functions with a negative $U'[3.6,7.3;0.92r]$. They  describe bound state. We note again that  this potential with $V_{\rm aadR}'(r)>0$ includes a tiny well located at $r=971$ [see (Right) Fig. 3]. After analyzing far-region and asymptotic wave functions, it turns out  that the tiny well could be  neglected effectively.

We observe the far-region and asymptotic wave functions corresponding to a local maximum potential [(Left)Fig. 5] with $V'_{\rm aadR}(r)<0$.
As is shown in (Left) Fig. 6, all functions are complex. Their real parts represent oscillating functions, while  Re$[\Psi^{\rm A}_{10}]$ represents a negative bound state.
This is so  because of $C(=-39)<-0.25$.
This  belongs to an unwanted case.

For an increasing  potential [(Left) Fig. 7] with $V'_{\rm aadR}(r)>0$, one has a half of a peak (quasibound state, $\Psi_4(r)\simeq\Psi^{\rm A}_4(r)$)[see (Right) Fig. 7].
As is shown in Fig. 8, the confluent hypergeometric function $U[-0.0008,10.6;10.5r]$ is constant nearly and thus, its first derivative  $U'[-0.0008,10.6;10.5r]$ is zero nearly ($a\simeq 0$).
This case represents a boundary between trapping well ($a<0$) and no trapping well ($a>0$).

According to the above prescription, the quasibound state  could be achieved when the first argument of $U[a,b;cr]$ is negative as
\begin{equation}
a<0 \to \quad \frac{B}{2A}>k+\frac{1}{2} \label{trap-well}
\end{equation}
which corresponds to the other condition to  possess  a  trapping well.
 On the other hand,
the bound state   could be found  when the first argument of $U[a,b;cr]$ is positive as
\begin{equation}
a>0 \to \quad \frac{B}{2A}<k+\frac{1}{2},\label{no-tw}
\end{equation}
which is regarded as the other condition for no trapping well.
The boundary case is obtained when $a$ is zero nearly,
\begin{equation}
a\simeq 0 \to \quad \frac{B}{2A}=k+\frac{1}{2}. \label{zero-tw}
\end{equation}

Finally, we obtain   two conditions for getting a trapping well as
\begin{equation}
M\mu^2> qQ_e\omega~[V'_{\rm aadR}(r)>0,~B>0]~{\rm and}~\frac{B}{2A}>k+\frac{1}{2}. \label{trap-well}
\end{equation}
The former condition  shows
\begin{equation}
\frac{M\mu}{qQ_e}>\frac{\omega}{\mu}, \label{trap-w2}
\end{equation}
which differs slightly  from the necessary condition for a trapping well  ($\frac{M\mu}{qQ_e}>1$) in ~\cite{Furuhashi:2004jk,Degollado:2013bha}.
This is so because of the bound state condition ($\frac{\omega}{\mu}<1$).
On the other hand, two conditions for no trapping well are given by
\begin{equation}
M\mu^2< qQ_e\omega~[V'_{\rm aadR}(r)<0,B<0]~{\rm and}~\frac{B}{2A}<k+\frac{1}{2}, \label{Notr-well}
\end{equation}
where the former condition could be written as~\cite{Myung:2022biw}
\begin{equation}
\frac{M\mu}{qQ_e} <\frac{\omega}{\mu} <1, \label{Notr-w2}
\end{equation}
which is different slightly from the condition for no trapping well as $\frac{M\mu}{qQ_e}<1$ appeared in Ref.\cite{Xu:2020fgq}.
\section{Geodesic analysis in the RN geometry}
In this section, we wish to explain how the known bounds of $\frac{M\mu}{qQ}>1( \frac{M\mu}{qQ}<1)$ arise from the pseudo-Newtonian  potential for a charged particle with mass $\mu$ and charge $q$ in the RN black holes with mass $M$ and charge $Q$~\cite{Ivanov:2005cf}. The RN black hole ($\hat{g}_{\mu\nu}$ and  $\hat{A}_\mu$)  is easily recovered from the dRN black hole (\ref{d-RN}) by resetting  $Q_e=Q$ and  $Q_m=0$.
We start with the Lagrangian for a charged particle moving in the RN geometry
\begin{equation}
{\cal L}=\frac{1}{2}\hat{g}_{\rm \mu\nu}\frac{dx^\mu}{d\lambda}\frac{dx^\nu}{d\lambda}-\frac{q}{\mu}\hat{A}_\mu \frac{dx^\mu}{d\lambda}
\end{equation}
with $\lambda=\tau/\mu$ the proper time per unit mass.
The radial equatorial ($\theta=\pi/2$) motion for a charged particle  takes the form
\begin{equation}
\Big(\frac{dr}{d\lambda}\Big)^2=E^2-V_{\rm g}(r),
\end{equation}
where $E=\partial {\cal L}/\partial \dot{t}$ is the conserved energy of the particle and $V_{\rm g}(r)$ is the the geodesic potential defined as
\begin{eqnarray}
V_{\rm g}(r)= \mu^2-\frac{2(M\mu^2-qQE)}{r} +\frac{L^2+Q^2(\mu^2-q^2)}{r^2}
-\frac{2ML^2}{r^3}+\frac{Q^2L^2}{r^4}, \label{g-pot}
\end{eqnarray}
where  $L=\partial {\cal L}/\partial \dot{\phi}$ is the conserved projection of the   particle's angular momentum on the axis of the black hole.
In association with $E\leftrightarrow\omega$, $Q \leftrightarrow Q_e$ with $Q_m=0$, and $L^2 \leftrightarrow l(l+1)$, the geodesic potential $V_{\rm g}(r)$ matches the scalar potential $\tilde{V}_{\rm dRN}(r)$ exactly
 when neglecting the terms from  introducing the tortoise coordinate $r_*$ in $V_{\rm dRN}(r)$.

 Focusing on a circular orbit ($dr/d\lambda=0$) in the equatorial plane, one finds two equations: $V_{\rm g}(r)=0 $ and
 $V'_{\rm g}(r)=0$. These  are viewed as a system of simultaneous equations from which one can determine $E$ and $L$ in terms of $M,Q,q$ and $\mu$.
 This is a process for eliminating $E$ and $L$.
 For $q/\mu \ll1$, one expands $E$ and $L$ as functions of $q/\mu$ with $\mu$ fixed: $E\simeq E_0(r)+qE_1(r) +{\cal O}(q/\mu)^2$ and $L\simeq L_0(r)+qL_1(r) +{\cal O}(q/\mu)^2$ with
 $E_0(r)$ the conserved energy for a neutral particle and $L_0(r)$ the conserved projection of the angular momentum  for a neutral particle.
 The centripetal force $F_0(r)$ acting on the neutral particle is given by $F_0(r)=[L_0^2(r)/E_0^2(r)]/r^3.$

 On the other hand, the centripetal force acting on the charged particle is given by
 \begin{equation}
 F(r)=\frac{L^2(r)}{E^2(r)}\frac{1}{r^3}=F_0(r)\Bigg(\frac{1+\frac{qL_1(r)}{\mu L_0(r)}}{1+\frac{qE_1(r)}{\mu E_0(r)}}\Bigg)^2\simeq F_0(r)\Big[1+2\frac{q}{\mu}\Big(\frac{L_1(r)}{L_0(r)}-\frac{E_1(r)}{E_0(r)}\Big)
 +{\cal O}\Big(\frac{q}{\mu}\Big)^2\Big].
 \end{equation}
 Its pseudo-Newtonian potential $V_{\rm M}(r)=-\int F(r)dr$ gives rise to the Manev potential of $V_{\rm M}(r)=G/r+H/r^2$ with $G$ and $H$ constants.
 A classical Manev potential to model a circular orbit of a lightly charged particle ($q/\mu \ll1$) can be  obtained  by seeking expansion over the powers of $M/r$.
 The leading and next-leading terms of this potential are given by
\begin{equation}
V_{\rm M}(r)\simeq \Big(1-\frac{qQ}{M\mu}\Big)\frac{M}{r}+\Big(2-\frac{Q^2}{2M^2}-\frac{9qQ}{4mM}\Big)\frac{M^2}{r^2}.
\end{equation}
Here, it is evident that the leading term is positive if $M \mu>qQ$ and  $q,Q>0$ (Coulomb repulsion).
This determines the centripetal force (due to the black hole) on the test charged particle.
Also, the leading term is negative if $M \mu<qQ$ and  $q,Q>0$. In this case, this determines the centrifugal force on the test charged particle.
This explains a direct appearance for the known bounds of $M \mu>qQ$ ($M \mu<qQ$) in the pseudo-Newtonian potential for a charged particle moving around the RN geometry.
\section{Summary and Discussions}
\begin{table*}[h]
\resizebox{16cm}{!}
{\begin{tabular}{|c|c|c|c|c|c|c|}
\hline
Case & $B$ &$a=k+\frac{1}{2}-\frac{B}{2A}$ &$r_{\rm e}=\frac{2C}{B}(C)$&trapping well&$qQ_m(\omega<\omega_c,q>\mu)$ \\ \hline
Fig. 1&$-0.027$  & 2.94 & $-323(4.4)$  &no  & 1 (yes,yes)   \\ \hline
Fig. 2&16.8& $-1.46$& 21.5(181)& yes (local minimum)&1 (no,yes)\\ \hline
Fig. 3&0.02& 3.63& 971(9.7)& no (a tiny well)&1 (yes,yes)\\ \hline
Fig. 5&$-0.16$& $2.2+6.2i$& 142($-39$)& no (local maximum)&5 (yes,yes)\\ \hline
Fig. 7&55.5& $-0.0008\simeq0$& 0.81(22.6)& no (boundary)&2 (yes,no)\\ \hline
\end{tabular}}
\caption{Results for asymptotic potential $V_{\rm aadR}(r)$ and far-region wave function analysis. All cases satisfy the condition for an asymptotic bound state ($\omega<\mu$).  $B>0(B<0)$ represents the case that the gravitational (electrostatic) force  exceeds the electrostatic (gravitational) force. $a<0(a>0$ denotes the condition for a quasibound (bound) state. $r_{\rm e}>r_+$ with $C>-0.25(C<-0.25)$ indicates the location of  a local minimum (maximum) far from the outer horizon.  $qQ_m$ denotes the quantization condition and $\omega<\omega_c$ represents the superradiance condition. Finally, $q>\mu$  may be  a necessary condition for superradiant instability when a cavity is placed around the dRN black hole.  }
\end{table*}
In this work, we have investigated  the superradiant (in)stability for dRN black holes by making use of  the  scalar potential $V_{\rm dRN}(r)$ (\ref{e-pot})  which differs by $q^2Q_m^2/r^2$ from the scalar potential $V_{\rm RN}(r)$ (\ref{RN-pot}) around the RN black hole. We stress  that an inclusion of  a magnetic charge $Q_m$ in $q^2Q_m^2/r^2$  did not induce a magnetic field like $\mu_{\rm eff} \propto B$ in the scalar potential $V_{\rm dRN}(r)$. This upsets  the original motivation of introducing dRN black holes.
We have obtained two conditions for getting a trapping well and for no trapping well  by analyzing all asymptotic potentials and far-region scalar wave functions.

We summarize the results  for  all asymptotic potential $V_{\rm aadR}(r)$ and far-region wave function analysis in Table 1.
One condition to have a  trapping well [$V'_{\rm aadR}(r)>0~(B>0)$] includes three potential types:  potential with a trapping well [(Left) Fig. 2], potential with  a tiny well [ (Left) Fig. 3], and  an increasing potential representing a boundary [(Left) Fig. 7]. Among these, the potential with a trapping well [(Left) Fig. 2] is satisfied with the other condition of $a<0(a=k+\frac{1}{2}-\frac{B}{2A})$.
The potential with a tiny well [ (Left) Fig. 3] goes with $a>0$ and  an increasing potential  [(Left) Fig. 7] has $a\simeq0$.
This implies  that two conditions for getting a trapping well are  given by $V'_{\rm aadR}(r)>0~(B>0)$ and $a<0$.
The former condition describes the Newton-Coulomb requirement for the gravitational force exceed the electrostatic force, while the latter is the condition to obtain
a quasibound state.
Here, it is important to point out that two conditions for a trapping well [(Left) Fig. 2]
and the superradiance condition of $\omega<\omega_c$ could not be satisfied simultaneously. This implies that the superradiant instability is not found naturally from a charged massive scalar propagating around the dRN black hole.  However, the superradiant instability of an electrically charged massive scalar  could be achieved  in certain parameter space if a cavity is placed around the dRN black hole~\cite{Herdeiro:2013pia,Degollado:2013bha,Hod:2013fvl,Hod:2016kpm}.

On the other hand, one condition for no trapping well [$V'_{\rm aadR}(r)<0~(B<0)$] involves two cases: potential without trapping well [(Left) Fig.1] and potential with  a local maximum [(Right) Fig. 5]. The potential without trapping well [(Left) Fig.1] is satisfied with the other condition of $a>0$.  The local maximum potential [(Right) Fig. 5] implies $a=2.2+6.2i$ (complex) and $C(=-39)<-0.25$. This means that two conditions for no trapping well are given by $V'_{\rm aadR}(r)<0~(B<0)$ and $a>0$. The former condition describes the Coulomb-Newton requirement for the electrostatic  force exceed the gravitational force, while the latter is the condition to find
a bound state.

Now, it is worth mentioning  that  one condition to have a trapping well Eq.(\ref{trap-w2}) differs slightly from the  known bound of $\frac{M\mu}{qQ}>1$~\cite{Furuhashi:2004jk,Degollado:2013bha}, while one condition for no trapping well Eq.(\ref{Notr-w2}) is different slightly from the known bound of $\frac{M\mu}{qQ}<1$~\cite{Xu:2020fgq}. We have explained  how these known bounds arise  from the pseudo-Newtonian  potential for a charged particle with mass $\mu$ and charge $q$ moving around  the RN black hole with mass $M$ and charge $Q$~\cite{Ivanov:2005cf}.

Finally, we  would like mention that the dRN black hole reduces to the RN black hole when replacing $Q_e^2+Q_m^2$ by $Q^2$.
Considering an electrically charged massive scalar perturbation around the RN black hole, one finds the corresponding scalar potential $V_{\rm RN}(r)$ in Eq.(\ref{RN-pot})~\cite{DiMenza:2014vpa,Benone:2015bst,Herdeiro:2013pia,Degollado:2013bha,Hod:2013fvl,Hod:2016kpm} which differs by $q^2Q^2_m/r^2$ from $V_{\rm dRN}(r)$  in Eq.(\ref{e-pot}). Even though we expect to obtain  similar results as in Table 1,  we have not found  such a complete  analysis for superradiant instability around the RN black holes in the literatures.

 \vspace{0.5cm}

{\bf Acknowledgments}
 \vspace{0.5cm}

This work was supported by a grant from Inje University for the Research in 2021 (20210040).

\end{document}